\documentclass[journal,twoside,web]{ieeecolor}

\usepackage{generic}

\usepackage[T1]{fontenc}
\usepackage[utf8]{inputenc}

\usepackage{amsmath,amssymb,amsfonts}
\usepackage{siunitx}
\usepackage{graphicx}
\usepackage{booktabs}
\usepackage{array}
\usepackage{tabularx}
\usepackage{makecell}
\usepackage{multirow}
\usepackage[caption=false,font=footnotesize]{subfig}
\usepackage{stfloats}

\usepackage{cite}
\usepackage{xurl}
\usepackage{textcomp}
\usepackage[hidelinks]{hyperref}

\usepackage[acronym]{glossaries}
\newacronym{amg}{AMG}{acceleromyography}
\newacronym{emg}{EMG}{electromyography}
\newacronym{rf}{RF}{Random Forest}

\graphicspath{{figures/}}

\def\BibTeX{{\rm B\kern-.05em{\sc i\kern-.025em b}\kern-.08em
T\kern-.1667em\lower.7ex\hbox{E}\kern-.125emX}}

\newcommand{\degunit}{^\circ}

\begin{document}

\title{Synchronized AMG and EMG Dataset of Lower-limb Muscle Activities in Everyday Training}




\author{Dongxu Tang, Shih Ying-Lei, Zhuoyi Ren, Jianting Liao, and Yitian Shao
\thanks{This work was supported by National Natural Science Foundation of China under Grant 62576119.}
\thanks{The authors are with the School of Computer Science and Technology, Harbin Institute of Technology, Shenzhen, Shenzhen, China (correspondence: shaoyitian@hit.edu.cn).}}

\maketitle

\thispagestyle{empty}

\begin{abstract}
Understanding how lower-limb muscle groups coordinate is important for studying movement impairment, rehabilitation, and physical performance. Reproducible analysis of this coordination requires multimodal recordings that relate local muscle-related signals with body-level kinematics. 
Complementing neural-level electrical activation captured by \gls{emg}, \gls{amg} provides a valuable mechanical approach to monitoring muscle activity.
Here, we introduce a synchronized, multimodal dataset for healthy-adult lower-limb activities. 
For data collection on the left leg, 16 triaxial accelerometers were evenly divided into four muscle-site clusters for \gls{amg} recording, complemented by four surface EMG channels. A 15-marker optical motion-capture (MoCap) system captured lower-body kinematics, with the resulting marker trajectories used to compute bilateral knee and ankle joint angles.
Our dataset contains 1,918 trials from 30 subjects across 16 task conditions.
We benchmark the dataset by estimating four joint angles from 300 ms windows of the {5--100\,Hz} band-pass-filtered \gls{amg} data and assess matched EMG features in a separate modality ablation.
In the primary cross subject benchmark, the four reference models achieved mean absolute errors of 8.840$^\circ$--9.591$^\circ$. The benchmark and ablation results characterize performance across subjects, tasks, and joint angles and examine the effects of sensor configuration, modality, the number of training subjects, and frequency representation.
The release includes documented timing definitions, processed data, and reproducible benchmark resources \url{https://dongxutang918-afk.github.io/SAME-Limb/}. 
\end{abstract}


\section{Introduction}
\label{sec:introduction}
\IEEEPARstart{C}{oordinated} activation of lower-limb muscles underlies locomotion, postural control, and functional movement. Characterizing this coordination is relevant to rehabilitation assessment, movement science, exercise, and sport, where wearable measurements can complement laboratory observations across repeated tasks \cite{Patel_2012,Porciuncula_2018,Campanini_2020}. 
While optical MoCap effectively quantifies segmental and whole-body kinematics, camera-based systems require calibrated laboratory space, suffer from line-of-sight constraints, and measure neither muscle electrical nor mechanical activity.
These limitations motivate the use of synchronized wearable measurements to complement MoCap systems \cite{Tao_2012,Patel_2012}.

Surface EMG measures electrical activity associated with neuromuscular activation. Mechanomyography (MMG) describes muscle-related mechanical oscillations and displacements transmitted to the body surface and is commonly regarded as a mechanical counterpart to EMG \cite{Ibitoye_2014,Islam_2013}. 
{Accelerometers are commonly used to record MMG \cite{Islam_2013,Ahn_2016}. Throughout this paper, \gls{amg} refers to the acceleration recorded at the selected muscle sites by the skin-mounted accelerometers.}
Small accelerometers offer a practical way to sample local mechanical variation at several positions around a muscle site and to pair these measurements with co-located EMG. 

Open datasets have made lower-limb sensing and biomechanics studies more comparable. Existing resources include bilateral neuromechanical locomotion recordings, lower-limb EMG with kinematic and kinetic references, multimodal wearable and MoCap datasets spanning cyclic and non-cyclic activities, high-density EMG with IMU and biomechanics measurements, and rehabilitation-oriented combinations of EMG, IMU, pressure, and MoCap \cite{Hu_2018,Lencioni_2019,Moreira_2021,Camargo_2021,Scherpereel_2023,Dimitrov_2023,Wei_2023,Schulte_2023,Jiang_2024,Boo_2025}. 
These resources address complementary questions and differ in 
task designs, sensor modalities, reference types, and sensor array configurations, as summarized in
Table~\ref{tab:related_dataset_landscape}.
However, a dataset combining a local AMG array with matched lower-limb EMG and synchronized optical MoCap joint angles is yet to be developed.
\begin{table*}[!t]
\caption{Representative lower-limb wearable and biomechanics datasets. The comparison summarizes dataset design rather than cross-dataset model performance. ``Local AMG array'' denotes multiple skin-mounted accelerometers arranged around matched muscle sites to record AMG; ``No'' does not imply that an IMU-based resource contains no accelerometers.}
\label{tab:related_dataset_landscape}
\centering
\scriptsize
\setlength{\tabcolsep}{2pt}
\renewcommand{\arraystretch}{1.06}
\begin{tabularx}{\textwidth}{
>{\raggedright\arraybackslash}p{0.18\textwidth}
c
>{\hsize=1.05\hsize\raggedright\arraybackslash}X
>{\hsize=1.20\hsize\raggedright\arraybackslash}X
>{\hsize=0.75\hsize\raggedright\arraybackslash}X
c}
\toprule
Resource &
No. subjects &
Tasks &
Wearable signals &
Reference labels &
Local AMG array \\
\midrule
Hu et al. \cite{Hu_2018} &
10 &
Unassisted locomotion and transitions &
Bilateral EMG and wearable kinematics &
Activity and transition labels &
No \\

Lencioni et al. \cite{Lencioni_2019} &
50 &
Walking and stairs &
Lower-limb EMG &
Kinematics and kinetics &
No \\

Moreira et al. \cite{Moreira_2021} &
16 &
Controlled-speed walking &
Lower-limb EMG &
Kinematics and kinetics &
No \\

Camargo et al. \cite{Camargo_2021} &
22 &
Level, ramps, stairs, and transitions &
IMU, EMG, and goniometers &
Kinematics and kinetics &
No \\

Scherpereel et al. \cite{Scherpereel_2023} &
12 &
Cyclic and non-cyclic tasks &
IMU and EMG &
MoCap kinematics and force-plate kinetics &
No \\

Dimitrov et al. \cite{Dimitrov_2023} &
10 &
Locomotion activities &
HD-sEMG and IMU &
Kinematics and kinetics &
No \\

Wei et al. \cite{Wei_2023} &
40 &
Lower-limb movements &
Surface EMG &
Kinematics and kinetics &
No \\

Schulte et al. \cite{Schulte_2023} &
55 &
Gait-related activities and transitions &
sEMG and IMUs &
Joint angles and wearable kinematics &
No \\

Jiang et al. \cite{Jiang_2024} &
24 &
Rehabilitation postures and gait &
EMG, IMU, insole pressure &
MoCap and task labels &
No \\

Boo et al. \cite{Boo_2025} &
120 &
Level, stair, and slope walking; sit-to-stand and stand-to-sit &
Surface EMG &
Full-body kinematics and kinetics &
No \\

This work &
30 &
Functional lower-limb tasks and treadmill walking &
Multi-site AMG and surface EMG &
MoCap-derived joint angles &
Yes \\
\bottomrule
\end{tabularx}
\end{table*}

In this work, we align muscle electrical and mechanical signals with body-level kinematic references. The dataset includes trial identifiers, data splits, window definitions, and benchmark baselines to facilitate result reproductions and future work expansions.
The resulting dataset is intended to complement, not replace, existing gait, rehabilitation, and multimodal activity datasets, and follows broader principles for reusable and well-documented research data \cite{Wilkinson_2016,Gebru_2021}.
Our dataset features lower-limb measurements from 30 healthy adults across 16 task conditions, yielding a total of 1,918 retained trials. 
Specifically, four muscle site accelerometer clusters comprise 16 triaxial accelerometers (48 axes) positioned around four matched EMG channels and provide the \gls{amg} recordings, while trajectories from 15 optical MoCap markers are used to compute bilateral knee and ankle joint angles.

To benchmark the dataset, we frame joint-angle estimation across four joints as a time-series regression task, evaluating a Random Forest baseline alongside three deep-learning models on \gls{amg} and \gls{emg} data.

\section{Dataset Overview}
The multimodal dataset is organized by subject, task, and trial. Each trial contains AMG, EMG, optical MoCap, time information, metadata, and joint angle labels derived from MoCap. Subject identifiers and task codes are defined in the data dictionary. Table~\ref{tab:dataset_inventory} summarizes the dataset size, task coverage, and trial completeness.

\begin{table}[!t]
\caption{Dataset inventory.}
\label{tab:dataset_inventory}
\centering
\footnotesize
\setlength{\tabcolsep}{3pt}
\renewcommand{\arraystretch}{1.08}
\begin{tabularx}{\columnwidth}{l >{\raggedright\arraybackslash}X}
\toprule
Item & Value \\
\midrule
Subjects & 30 healthy adults; 15 female and 15 male \\
Demographics & Age 23.4\(\pm\)2.1 years; height 169.2\(\pm\)8.1 cm; weight 60.8\(\pm\)11.6 kg; BMI 21.1\(\pm\)2.7 kg/m\(^2\) \\
Tasks & 16 conditions: 11 functional actions and 5 walking speeds \\
Trials & 1,918 retained trials from 1,920 expected \\
Wearables & 16 triaxial accelerometers (48 axes) for AMG and 4 matched EMG channels at four muscle sites \\
Labels & Joint angle labels derived from MoCap \\
\bottomrule
\end{tabularx}
\end{table}
The subjects were 21--33 years old, with heights of 155--188 cm, weights of 42--89 kg, and BMI values of 16.1--29.1 kg/m$^2$.
Each subject performed 11 functional lower-limb actions and five treadmill-walking conditions. The functional tasks were 
{\textit{Quiet Stand}, \textit{Quiet Sit}, \textit{Deadlift}, \textit{Deep Squat}, \textit{Stair Ascent}, \textit{Left Lunge}, \textit{Single-leg Stance}, \textit{Forward Lunge}, \textit{Vertical Jump}, \textit{Stand-sit 
transition}, and \textit{Heel Raise}. The walking conditions were \textit{Walk at 1\,km/h}, \textit{Walk at 2\,km/h}, \textit{Walk at 3\,km/h}, \textit{Walk at 4\,km/h}, and \textit{Walk at\,5 km/h}. Each subject was expected to complete four trials for each condition. The protocol therefore included 1,920 expected trials, of which 1,918 were retained, since two trials of \textit{Walk at 2\,km/h} from one subject were excluded due to data loss from a malfunctioning accelerometer. The identifiers for these excluded trials are documented in the data dictionary.

Figure~\ref{fig:dataset_overview} summarizes the sensor layout, task examples, and representative signals. AMG and EMG were recorded at four sites on the left leg: 
{the rectus femoris (RF), vastus medialis (VM), tibialis anterior (TA), and gastrocnemius (GA). }
At each site, one EMG sensor was paired with four miniature triaxial accelerometers. Fifteen optical MoCap markers were used to calculate the joint angle labels. Device models, sampling rates, and acquisition hardware are described in Methods.


\begin{figure*}[!t]
\centering
\includegraphics[width=\textwidth]{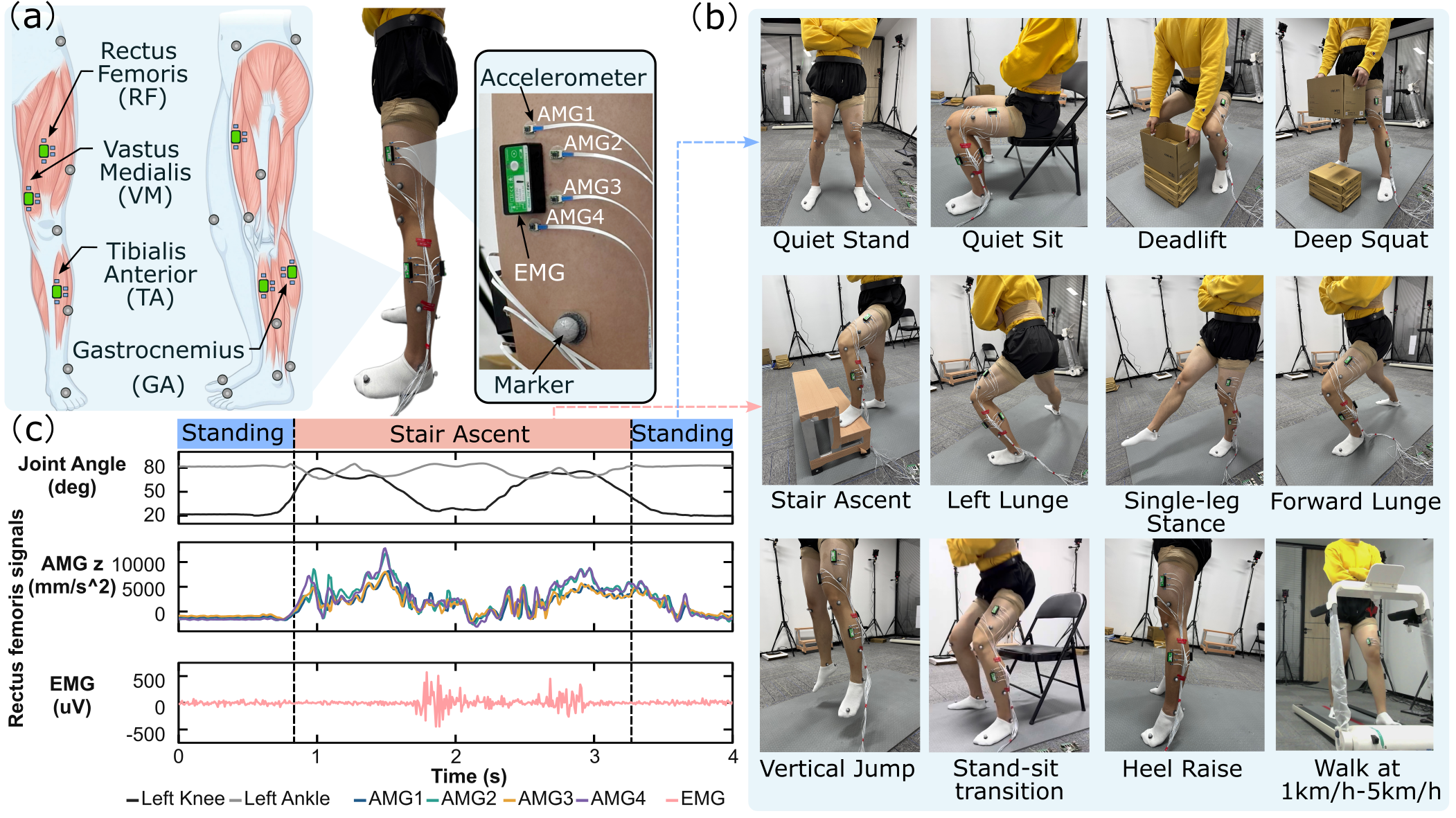}
\caption{Dataset overview, sensor placement, task examples, and representative multimodal signals. (a) Surface EMG sensors were placed over the rectus femoris, vastus medialis, tibialis anterior, and gastrocnemius. Four miniature triaxial accelerometers were arranged around each EMG site, and optical MoCap markers were placed at lower limb landmarks. (b) The protocol included 11 functional lower limb actions and treadmill walking at 1--5 km/h. (c) One \textit{Stair Ascent} trial shows the left knee and left ankle joint angles derived from MoCap, four AMG signals from the accelerometers' \(z\) axes at the rectus femoris site, and the matched raw EMG signal on the same time axis.
}
\label{fig:dataset_overview}
\end{figure*} 

\begin{figure*}[!t]
\centering
\includegraphics[width=\textwidth]{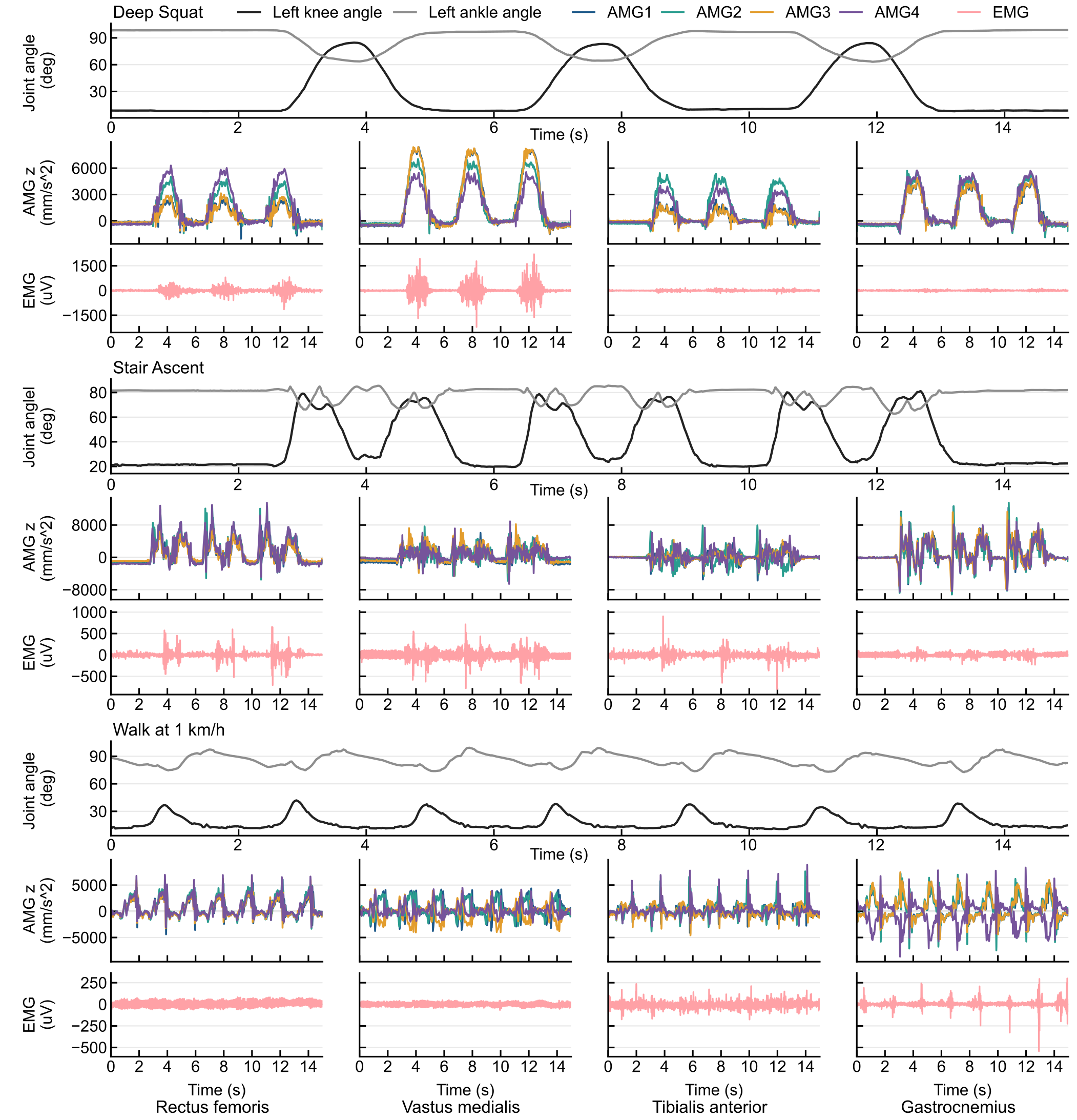}
\caption{Representative multimodal signals from three recorded trials. The rows show \textit{Deep Squat}, \textit{Stair Ascent}, and \textit{Walk at 1\,km/h}. In each row, the MoCap plot shows the left knee and left ankle joint angles. The four muscle site columns show AMG signals from the accelerometers' \(z\) axes and the matched raw EMG signal at the RF, VM, TA, and GA sites. All signal plots in a row use the same time axis.}
\label{fig:aligned_signals}
\end{figure*}
\section{Methods}

\subsection{Acquisition Protocol}
The study protocol was approved by the Medical Ethics Committee of Harbin Institute of Technology (Approval No. HIT-2024046; July 8, 2024). All subjects provided written informed consent before data collection.
Each recording session captured lower-limb wearable signals and optical MoCap during the full task protocol. Surface EMG was recorded using a DataLITE WS450 wireless acquisition system with four surface EMG sensors (LE230, Biometrics Ltd.) positioned over the rectus femoris, vastus medialis, tibialis anterior, and gastrocnemius. At each EMG site, four custom triaxial accelerometers (AIS2IHTR, STMicroelectronics) formed a muscle-site \gls{amg} cluster. Each cluster was connected to a Raspberry Pi 5, and four clusters operated in parallel. This arrangement preserves a direct mapping among \gls{amg} clusters, the anatomical site, the matched EMG channel, and the released metadata.
{The total cost of our customized 16-sensor AMG acquisition system is \$300 while the cost of 4-sensor EMG system is more than \$10000.}

The accelerometer channels used for AMG were sampled at 1600 Hz, corresponding to a Nyquist frequency of 800 Hz. 
Raw accelerometer measurements are expressed in units of {g/digit} ($1\text{g} = 9.80665\,\text{m/s}^2$). To convert these raw values to $\text{mm/s}^2$, they are multiplied by the scale factor $a = 19.14258$.
EMG was sampled at 1000 Hz and stored as voltage measurements \(\mu\mathrm{V}\).
Fifteen reflective markers were recorded at 90 Hz using 16 infrared motion capture cameras (Mars2H, NOKOV) controlled using the accompanied software (XINGYING 3.4.0.3957, NOKOV). The marker trajectories were used to calculate the joint angles of both knees and ankles.

Heart rate was monitored using a smart bracelet (Band 9, Huawei) as a safety and recovery indicator. Recording was paused when the displayed heart rate was elevated relative to the subject's resting or recovery state, or when the subject reported fatigue or discomfort. Recording resumed after rest and confirmation that the subject was ready to continue. However, heart rate data was not recorded and not presented in the released dataset.

{For each subject and task condition, four trials were recorded and labeled \textit{A1}, \textit{A2}, \textit{A3}, and \textit{A4}. Except for treadmill walking and the static \textit{Quiet Stand}, \textit{Quiet Sit} and \textit{Single-leg Stance} tasks, three auditory cues at approximately 2, 6, and 10 s were used in each trial to guide movement timing. Auditory cue timing is included in the released dataset. Further details of cue timing, task instructions, treadmill preparation, and rest criteria are provided in Supplementary Information (SI) Sections~S4--S7 and Table~S2.}

\subsection{Processed Data Organization}
Each retained trial is stored in one compressed data file (python NPZ) containing wearable signals, MoCap arrays, a time vector for each modality, metadata, and joint angle arrays. Exact array names, dimensions, units, and metadata fields are listed in the data dictionary. Figure~\ref{fig:aligned_signals} shows representative multimodal signals from three trials.
For \gls{amg}, each file contains three aligned arrays: recorded device values, DC corrected values, and values after a \textit{5\,Hz} high-pass filter. For each aligned \gls{amg} channel, DC correction was performed by subtracting the mean of the initial \SI{2}{s} from all samples in the same trial. The rule for shorter trials is given in SI Section~S9.1. The primary benchmark input was generated by applying a zero phase Butterworth \textit{5--100\,Hz} band-pass filter to the DC corrected signals from each complete trial before window extraction. The resulting input retained all 48 \gls{amg} channels.
For \gls{emg}, each file contains four raw channels and the corresponding signals after a zero phase Butterworth \textit{20--450\,Hz} band pass filter. 

Optical MoCap marker and skeleton data were parsed into named arrays before the joint angles were calculated. When duplicate marker or skeleton groups were present, the valid group was retained. The selection rules and array definitions are documented in the data dictionary so that the joint angle labels can be reconstructed.

\subsection{Synchronization of Multimodal Data} %

Synchronization was performed in two stages. First, the four \gls{amg} acquisition streams were mapped to a common time grid. Second, \gls{amg}, EMG, and MoCap were cropped to their shared time range. The four accelerometers at each muscle site shared one file time axis, so the 16 accelerometers formed four \gls{amg} acquisition streams.

Let \(\mathcal{P}\) denote the four \gls{amg} acquisition streams. For stream \(p\in\mathcal{P}\), let \(t_{S,\text{AMG}}^{(p)}\) and \(t_{F,\text{AMG}}^{(p)}\) denote its recorded start and final sample time, respectively. 
For a stream containing \(N_p\) samples,
\begin{align}
t_{F,\text{AMG}}^{(p)}
=
t_{S,\text{AMG}}^{(p)}
+
\frac{N_p-1}{f_{\text{AMG}}},
\end{align}
with $f_{\text{AMG}}=1600~\mathrm{Hz}$.
The common AMG start and end times were defined as
$t_{S,\text{AMG}}
=
\max_{p\in\mathcal{P}}
t_{S,\text{AMG}}^{(p)}$ and 
$t_{F,\text{AMG}}
=
\min_{p\in\mathcal{P}}
t_{F,\text{AMG}}^{(p)}$.
The four AMG streams were linearly interpolated onto the common 1600 Hz time grid between \(t_{S,\text{AMG}}\) and \(t_{F,\text{AMG}}\). Values outside the recorded range were not extrapolated.

Let \(t_{S,\text{EMG}}\) and \(t_{F,\text{EMG}}\) denote the first and final EMG sample times, respectively. For \(N_{\text{EMG}}\) samples,
\begin{align}
t_{F,\text{EMG}}
=
t_{S,\text{EMG}}
+
\frac{N_{\text{EMG}}-1}{f_{\text{EMG}}},
\end{align}
where $f_{\text{EMG}}=1000~\mathrm{Hz}$.
MoCap retained its native timestamps after metadata parsing. The first and last valid MoCap timestamps are denoted by \(t_{S,\text{MoCap}}\) and \(t_{F,\text{MoCap}}\), respectively.

The shared time range of AMG, EMG, and MoCap was defined as
\begin{align}
t_{S}
&=
\max\left(
t_{S,\text{AMG}},
t_{S,\text{EMG}},
t_{S,\text{MoCap}}
\right),\\
t_{F}
&=
\min\left(
t_{F,\text{AMG}},
t_{F,\text{EMG}},
t_{F,\text{MoCap}}
\right).
\end{align}
The three modalities were cropped to this shared time range.



\subsection{Joint Angle Construction}
Four joint angles derived from MoCap serve as the regression targets. They are the right knee, left knee, right ankle, and left ankle angles. The angles are calculated from segment vectors defined by the markers. For any two nonzero segment vectors \(\mathbf{u}\) and \(\mathbf{v}\), the angle between them is
\begin{equation}
\phi(\mathbf{u},\mathbf{v}) =
\frac{180}{\pi}\arccos\left(
\operatorname{clip}\left[
\frac{\mathbf{u}^{\mathsf{T}}\mathbf{v}}
{\lVert \mathbf{u} \rVert_2 \lVert \mathbf{v} \rVert_2},
-1, 1\right]
\right).
\label{eq:segment_angle}
\end{equation}
The clipping operation prevents numerical round-off from moving the cosine argument outside the valid arccosine domain.
The knee angles are calculated from the thigh and shank vectors as
\begin{equation}
y_{\mathrm{knee}} = 180^\circ - \phi(\mathbf{u}_{\mathrm{thigh}},\mathbf{u}_{\mathrm{shank}}).
\label{eq:knee_joint_angle}
\end{equation}
For the right knee, \(\mathbf{u}_{\mathrm{thigh}}\) points from R.Knee to R.Thigh and \(\mathbf{u}_{\mathrm{shank}}\) points from R.Knee to R.Ankle. The left knee uses the corresponding markers on the left side. For the ankle angles, the interior angle is retained:
\begin{equation}
y_{\mathrm{ankle}} = \phi(\mathbf{u}_{\mathrm{shank}},\mathbf{u}_{\mathrm{foot}}),
\label{eq:ankle_joint_angle}
\end{equation}
where \(\mathbf{u}_{\mathrm{shank}}\) points from the ankle marker toward the shank marker and \(\mathbf{u}_{\mathrm{foot}}\) points from the ankle marker toward the toe marker. Frozen marker indices and invalid-data rules are provided in SI Section~S14 and SI Table~S6. Note that these marker-derived joint angles may differ slightly from the true anatomical kinematics of the human body, due to variations in anthropometric dimensions and marker placement.


\section{Dataset Benchmark}
\label{sec:dataset_benchmark}

This benchmark evaluates how much information about lower limb movement can be decoded from AMG. The four joint angles derived from MoCap are used as reference targets. The first group of experiments evaluates the overall estimation performance and its variation across subjects, tasks, and joint angles. The second group examines the effects of frequency content, modality, sensor placement, sensor density, and the number of training subjects.

\subsection{Benchmark Definition and Common Settings}
\label{subsec:benchmark_common_settings}


\subsubsection{Benchmark Task and Window Construction}

The benchmark estimates four joint angles from short AMG windows. Dividing each trial into windows gives all reference models the same input length and produces estimates at regular time intervals. Each window is treated as one model input. Because adjacent windows overlap, they are not treated as independent statistical observations.

Let \(\mathbf{x}[n]\in\mathbb{R}^{48}\) denote the 48 AMG channel values at sample \(n\). For window \(i\), the input is
\begin{equation}
\mathbf{X}_i=
\left[
\mathbf{x}[n_i],
\mathbf{x}[n_i+1],
\ldots,
\mathbf{x}[n_i+479]
\right]
\in\mathbb{R}^{48\times480},
\label{eq:window_end_protocol}
\end{equation}
where \(n_i\) is the first sample index of the window. Each window contains 480 samples and spans 300 ms at 1600 Hz. Window endpoints are spaced by 100 ms, producing estimates at 10 Hz. Let \(t_i^{\mathrm{end}}\) denote the end time of window \(i\) on the common AMG time grid. The exact sample bounds and endpoint times are stored in the released window manifests.


\subsubsection{Target Settings and Label Assignment}
\label{subsubsec:target_settings}

Let \(h\in\{1,2,3\}\) index the three target settings. Each setting specifies the time of the joint angle label relative to the end of an AMG window:
\begin{enumerate}
\item
\textit{Window-end estimation} (\(h=1\)): Predicting the state at the exact end of the window, with \(\delta_1=0\). This setting is used for the primary benchmark.

\item
\textit{Centered-window offline estimation} (\(h=2\)): Predicting the state 150 ms prior to the window end, with \(\delta_2=-0.15~\mathrm{s}\).

\item
\textit{Forecasting} (\(h=3\)): Predicting the state 100 ms after the window end, with \(\delta_3=0.10~\mathrm{s}\).
\end{enumerate}

For window \(i\), the elapsed time from the common AMG start to the window end is
$
\tau_i
=
t_i^{\mathrm{end}}
-
t_{S,\text{AMG}}.
$
For setting \(h\), the corresponding time on the native MoCap time axis is
\begin{equation}
q_{i,h}
=
t_{S,\text{MoCap}}
+
\tau_i
+
\delta_h,
\label{eq:clean_relative_clock}
\end{equation}
where \(\delta_h\) is the time shift defined above.

Because MoCap is recorded as discrete frames, the MoCap frame nearest to \(q_{i,h}\) is selected:
\begin{equation}
n_{i,h}
=
\arg\min_n
\left|
t_{n,\text{MoCap}}
-
q_{i,h}
\right|,
\label{eq:nearest_native_frame}
\end{equation}
where \(t_{n,\text{MoCap}}\) is the native timestamp of MoCap frame \(n\), and \(n_{i,h}\) is the index of the nearest MoCap frame. 
Windows with \(\left|t_{n_{i,h},\text{MoCap}}-q_{i,h}\right|>1/90~\mathrm{s}\)
are deemed as invalid data segment and thereby excluded form our benchmark tests.

Before joint angle label validity was checked, the benchmark contained 264,081 windows from 1,918 retained trials. Each window contained 480 samples from 48 AMG channels and 300 native samples from four EMG channels. For the primary \textit{Window-end estimation}, 263,999 windows had valid joint angle labels; the within subject and cross subject test sets contained 65,919 and 61,687 valid windows, respectively. Complete window counts for all three settings and data splits are provided in SI Table~S3.

\subsubsection{Data Splits}
\label{subsubsec:data_splits}

Within subject evaluation used \textit{A1} and \textit{A2} for training, \textit{A3} for validation, and \textit{A4} for testing. Trial assignment was completed before window extraction. Cross subject evaluation assigned 20 subjects to training, 3 subjects to validation, and 7 subjects to testing. All trials from one subject remained in the same cross subject partition.

\subsubsection{Input Representations and Reference Models}

The time series models use the full sequence \(\mathbf{X}_i\). For \textit{Random Forest}, eight features are calculated separately for each AMG channel.

Let \(x_{ic,n}\) denote sample \(n\) from AMG channel \(c\) in window \(i\), where \(c=1,\ldots,48\), \(n=1,\ldots,L\), and \(L=480\). The eight features calculated for each AMG channel are defined in Table~\ref{tab:amg_feature_definitions}.

\begin{table}[!t]
\caption{AMG features calculated for channel \(c\) in window \(i\), where \(x_{ic,n}\) denotes sample \(n\) and \(L=480\).}
\label{tab:amg_feature_definitions}
\centering
\footnotesize
\setlength{\tabcolsep}{2pt}
\renewcommand{\arraystretch}{1.08}
\begin{tabularx}{\columnwidth}{
@{}
>{\raggedright\arraybackslash}p{0.33\columnwidth}
>{\centering\arraybackslash}X
@{}}
\toprule
Feature & Definition \\
\midrule

Mean &
\(\displaystyle
\mu_{ic}
=
\frac{1}{L}
\sum_{n=1}^{L}x_{ic,n}
\) \\

Population standard deviation &
\(\displaystyle
\sigma_{ic}
=
\left[
\frac{1}{L}
\sum_{n=1}^{L}
(x_{ic,n}-\mu_{ic})^2
\right]^{1/2}
\) \\

Root mean square &
\(\displaystyle
\rho_{ic}
=
\left[
\frac{1}{L}
\sum_{n=1}^{L}
x_{ic,n}^{2}
\right]^{1/2}
\) \\

Minimum &
\(\displaystyle
x^{\min}_{ic}
=
\min_{1\leq n\leq L}x_{ic,n}
\) \\

Maximum &
\(\displaystyle
x^{\max}_{ic}
=
\max_{1\leq n\leq L}x_{ic,n}
\) \\

Range &
\(\displaystyle
\delta_{ic}
=
x^{\max}_{ic}-x^{\min}_{ic}
\) \\

Mean absolute deviation &
\(\displaystyle
\eta_{ic}
=
\frac{1}{L}
\sum_{n=1}^{L}
|x_{ic,n}-\mu_{ic}|
\) \\

Sum of squares &
\(\displaystyle
q_{ic}
=
\sum_{n=1}^{L}x_{ic,n}^{2}
\) \\

\bottomrule
\end{tabularx}
\end{table}

The eight values, in the order listed in Table~\ref{tab:amg_feature_definitions}, form \(\mathbf{f}_{ic}\in\mathbb{R}^{8}\). Concatenating the features from all 48 AMG channels gives a feature vector with 384 dimensions. The range is derived from the within-window minimum and maximum of the same channel, and \(q_{ic}=L\rho_{ic}^{2}\). These two related features were retained because all reported experiments using AMG features used the same set of eight features.

For EMG, let \(e_{id,m}\) denote sample \(m\) from EMG channel \(d\) in window \(i\), where \(d=1,\ldots,4\), \(m=1,\ldots,M\), and \(M=300\). For each channel, we calculated mean absolute value (MAV), waveform length (WL), zero crossings (ZC), and slope sign changes (SSC). These four features form the widely used Hudgins time domain feature set \cite{Hudgins_1993,Phinyomark_2018}. MAV summarizes signal magnitude, WL summarizes accumulated sample changes, ZC counts sign changes, and SSC counts changes in local slope direction.

For compact notation, define
\(
\Delta^{-}e_{id,m}
=
e_{id,m}-e_{id,m-1},
\qquad
\Delta^{+}e_{id,m}
=
e_{id,m}-e_{id,m+1}.
\)
The four EMG features are defined in Table~\ref{tab:emg_feature_definitions}.

\begin{table}[!t]
\caption{EMG features calculated for channel \(d\) in window \(i\) \protect\cite{Hudgins_1993,Phinyomark_2018}. Here,\(\mathbb{I}[\cdot]\) equals one when its condition is satisfied and zero otherwise; \(\land\) denotes logical AND; and \(\epsilon=0\). Strict inequalities are used for ZC and SSC.}
\label{tab:emg_feature_definitions}
\centering
\footnotesize
\setlength{\tabcolsep}{2pt}
\renewcommand{\arraystretch}{1.10}
\begin{tabularx}{\columnwidth}{
@{}
>{\raggedright\arraybackslash}p{0.28\columnwidth}
>{\centering\arraybackslash}X
@{}}
\toprule
Feature & Definition \\
\midrule

Mean absolute value (MAV) &
\(\displaystyle
\mathrm{MAV}_{id}
=
\frac{1}{M}
\sum_{m=1}^{M}
|e_{id,m}|
\) \\

Waveform length (WL) &
\(\displaystyle
\mathrm{WL}_{id}
=
\sum_{m=2}^{M}
|\Delta^{-}e_{id,m}|
\) \\

Zero crossings (ZC) &
\(\displaystyle
\mathrm{ZC}_{id}
=
\sum_{m=2}^{M}
\mathbb{I}\!\left[
\substack{
e_{id,m-1}e_{id,m}<0\\
{}\land |\Delta^{-}e_{id,m}|>\epsilon
}
\right]
\) \\

Slope sign changes (SSC) &
\(\displaystyle
\mathrm{SSC}_{id}
=
\sum_{m=2}^{M-1}
\mathbb{I}\!\left[
\substack{
\Delta^{-}e_{id,m}\Delta^{+}e_{id,m}>0\\
{}\land |\Delta^{-}e_{id,m}|>\epsilon\\
{}\land |\Delta^{+}e_{id,m}|>\epsilon
}
\right]
\) \\

\bottomrule
\end{tabularx}
\end{table}

We used \(\epsilon=0\) and strict inequalities for ZC and SSC. MAV differs from the AMG mean absolute deviation in Table~\ref{tab:amg_feature_definitions}: MAV is calculated relative to zero, whereas the AMG mean absolute deviation is calculated relative to the mean of the AMG channel. The four features from four EMG channels give 16 EMG features. Combining the 384 AMG features with the 16 EMG features gives a feature vector with 400 dimensions.

{Utilizing the features defined above, machine learning models were used to predict the lower-limb joint angles.}
Random Forest was used as a nonlinear reference model for the AMG features because it can model nonlinear relations while requiring only a small number of main settings \cite{Breiman_2001}. The model used 100 trees, a maximum depth of 15.

The AMG sequence \(\mathbf{X}_i\) was also evaluated using three time series models implemented with tsai version 1.0.1 \cite{tsai}:
\begin{enumerate}
\item
\textit{TCN}: Eight temporal convolution layers with 25 channels per layer, kernel size 7, and 76,154 trainable parameters \cite{Bai_2018}.

\item
\textit{LSTMPlus}: One unidirectional LSTM layer with hidden size 100, followed by a regression head using the final time step, with 60,404 trainable parameters \cite{Hochreiter_1997}.

\item
\textit{InceptionTimePlus}: Residual depth 6, 32 filters, kernel size 40, bottleneck layers, and 464,388 trainable parameters \cite{Fawaz_2020}.
\end{enumerate}

Each reported model configuration was trained once. For the time series models, normalization statistics for each input channel and each output were calculated from the training partition only. Training used mean squared error loss and AdamW, and the checkpoint with the lowest validation MAE was retained. Complete training settings, including the random seed, are provided in SI Sections~S11.2--S11.3 and Table~S5.

\subsubsection{Evaluation Metrics and Statistical Reporting}

Model performance is evaluated using mean absolute error (MAE), root mean squared error (RMSE), and Pearson \(r\). Let \(N\) denote the number of valid windows used to calculate a metric. For the primary cross subject \textit{Window-end estimation}, \(N=61{,}687\). The values for the other target settings and data splits are provided in SI Table~S3. For subject, task, and ablation analyses, \(N\) denotes the number of valid windows in the corresponding subset. Let \(K=4\) denote the number of joint angles.

For window \(i\) and joint angle \(k\), \(y_{ik}\) is the reference value and \(\hat y_{ik}\) is the predicted value. Their means over the \(N\) valid windows are
\begin{equation}
\bar y_k
=
\frac{1}{N}
\sum_{i=1}^{N}y_{ik},
\qquad
\bar{\hat y}_k
=
\frac{1}{N}
\sum_{i=1}^{N}\hat y_{ik}.
\end{equation}
The reported metrics are

\begin{equation}
\mathrm{MAE}
=
\frac{1}{K}
\sum_{k=1}^{K}
\frac{1}{N}
\sum_{i=1}^{N}
\left|
\hat y_{ik}-y_{ik}
\right|.
\label{eq:mae}
\end{equation}

\begin{equation}
\mathrm{RMSE}
=
\frac{1}{K}
\sum_{k=1}^{K}
\sqrt{
\frac{1}{N}
\sum_{i=1}^{N}
(\hat y_{ik}-y_{ik})^2
}.
\label{eq:rmse}
\end{equation}

\begin{equation}
r
=
\frac{1}{K}
\sum_{k=1}^{K}
\frac{
\sum_{i=1}^{N}
(y_{ik}-\bar y_k)
(\hat y_{ik}-\bar{\hat y}_k)
}{
\sqrt{
\sum_{i=1}^{N}
(y_{ik}-\bar y_k)^2
}
\sqrt{
\sum_{i=1}^{N}
(\hat y_{ik}-\bar{\hat y}_k)^2
}
}.
\label{eq:pearson_r}
\end{equation}


{For the primary benchmark, each metric is first calculated separately for the four joint angles over all valid cross subject test windows. The reported value is then obtained by averaging the four joint angle values.

For cross-subject analyses, the metric is calculated separately for each test subject, so every subject contributes one value. For task analysis, MAE is first averaged over the valid windows and four joint angles for each subject and task, and these subject values are then averaged across the seven test subjects. For joint angle analysis, MAE is calculated separately for each subject and each joint angle. The ablation analyses also calculate one MAE for each test subject before averaging across subjects.}

\subsection{Benchmark Evaluation}
\label{subsec:benchmark_evaluation}

\subsubsection{Cross-subject Benchmark} 
\label{subsubsec:primary_cross_subject}
This experiment evaluates whether the \gls{amg} input can estimate joint angles for subjects excluded from model training. It uses the cross-subject split and \textit{Window-end estimation}. The \gls{amg} data was band-pass filtered between 5 and 100\,Hz, to eliminate the impact of body kinematics, preserving only the skin dynamics.

\begin{table}[!t]
\caption{Primary cross subject results for \textit{Window-end estimation}.}
\label{tab:primary_benchmark_results}
\centering
\footnotesize
\renewcommand{\arraystretch}{1.08}
\setlength{\tabcolsep}{2pt}

\begin{tabular*}{\columnwidth}{
@{\extracolsep{\fill}}lrrr@{}}
\toprule
Model &
MAE (\(\degunit\)) &
RMSE (\(\degunit\)) &
Pearson \(r\) \\
\midrule
\textit{Random Forest}     & 9.591 & 13.639 & 0.634 \\
\textit{TCN}               & 8.840 & 12.515 & 0.672 \\
\textit{LSTMPlus}          & 9.535 & 13.589 & 0.630 \\
\textit{InceptionTimePlus} & 9.426 & 12.932 & 0.658 \\
\bottomrule
\end{tabular*}
\end{table}

The results show that the \textit{5--100\,Hz} \gls{amg} input contains information related to the four joint angles under cross subject evaluation. Among the four reference models, \textit{TCN} produced the lowest MAE and RMSE and the highest Pearson \(r\). Differences across subjects, tasks, and joint angles are examined below.

\subsubsection{Subject, Task, and Joint Angle Analysis}
\label{subsubsec:subject_task_joint}

To examine whether the overall results were consistent across the test data, performance was also summarized by subject, task, and joint angle.
\begin{figure*}[!t]
\centering
\includegraphics[width=\textwidth]{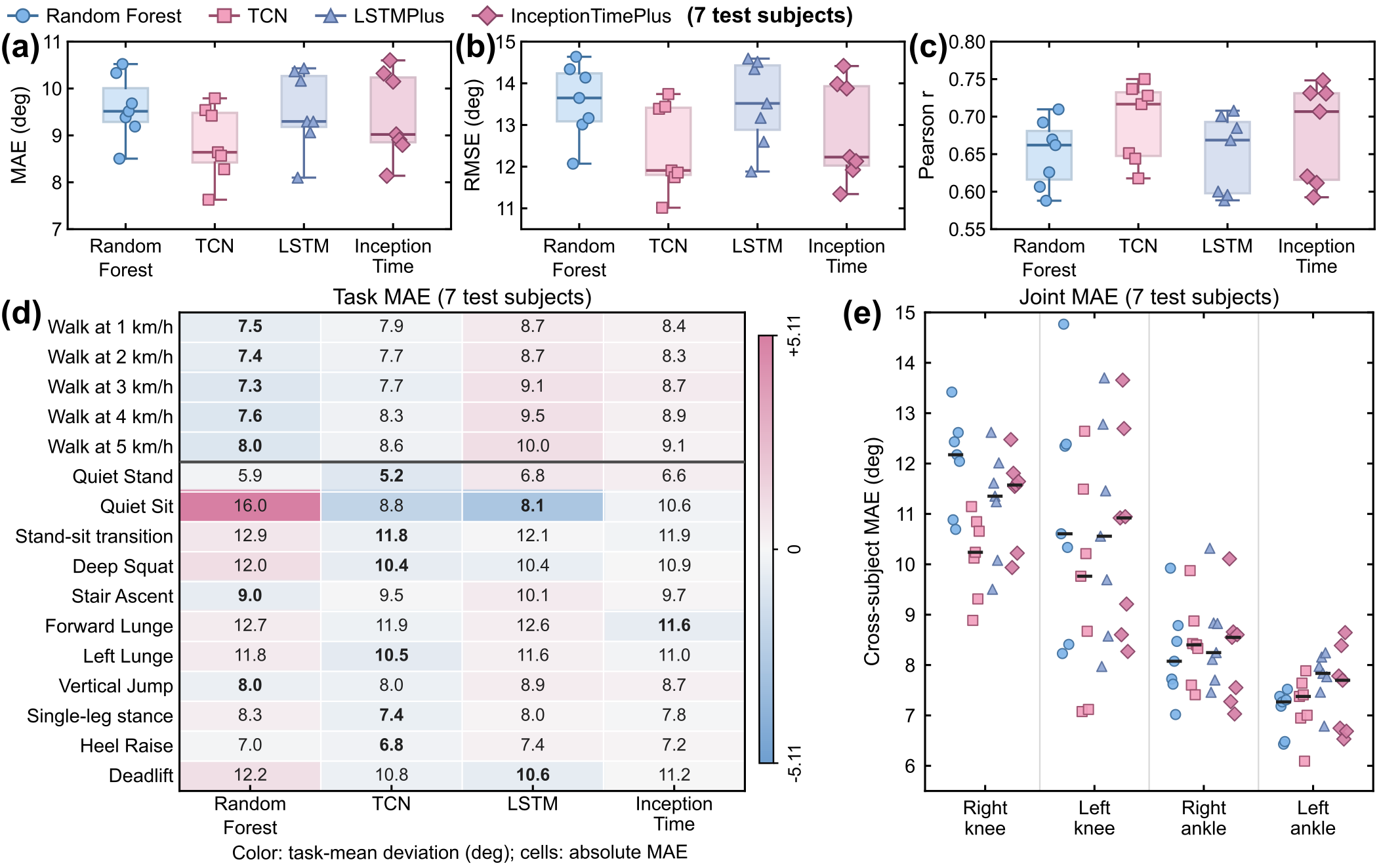}
\caption{Cross subject results for \textit{Window-end estimation} using \textit{Random Forest}, \textit{TCN}, \textit{LSTMPlus}, and \textit{InceptionTimePlus}. Panels (a)--(c) show MAE, RMSE, and Pearson \(r\) across the seven test subjects. Each point represents one subject; boxes show the interquartile range, center lines show medians, and whiskers show the observed range. (d) Task MAE was calculated separately for each subject and then averaged equally across the seven test subjects. Bold values show the lowest MAE for each task, and cell color shows deviation from the four-model mean for that task. (e) Points show joint angle MAE for each test subject; black ticks show the median.}
\label{fig:benchmark_stats}
\end{figure*}
Figure~\ref{fig:benchmark_stats} shows clear variation across test subjects, tasks, and joint angles. The five walking conditions and \textit{Quiet Stand} generally produced lower errors, whereas \textit{Quiet Sit}, \textit{Stand-sit transition}, \textit{Deep Squat}, the lunge tasks, and \textit{Deadlift} produced larger errors for at least some models. \textit{Quiet Sit} showed particularly large variation across models. Across the four reference models, knee angle errors were generally higher than ankle angle errors. These results show that one overall metric does not fully describe the difficulty of the benchmark.

\subsubsection{Representative Temporal Predictions}
\label{subsubsec:temporal_predictions}

Figure~\ref{fig:representative_prediction_trace} shows one cross subject test trial from each of three tasks. These examples show how prediction errors change over time.
\begin{figure*}[!t]
\centering
\includegraphics[width=\textwidth]{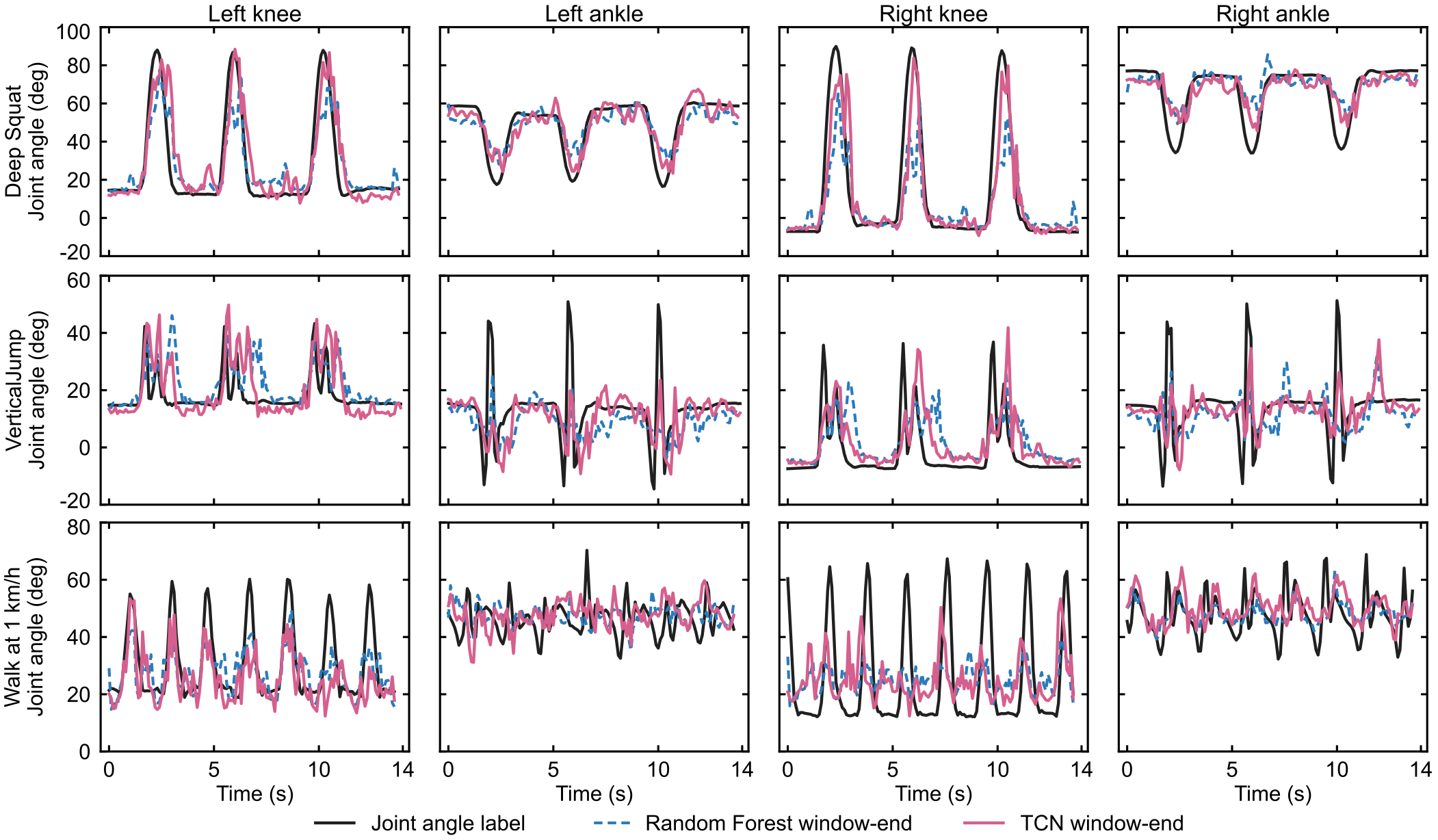}
\caption{Representative cross subject predictions for \textit{Window-end estimation}. Rows show trials DS-01 (\textit{Deep Squat}), VJ-06 (\textit{Vertical Jump}), and WK-02 (\textit{Walk at 1\,km/h}). Columns show the left knee, left ankle, right knee, and right ankle. Black curves show the joint angle labels, blue dashed curves show \textit{Random Forest} predictions, and pink curves show \textit{TCN} predictions.}
\label{fig:representative_prediction_trace}
\end{figure*}
The predictions followed the main movement patterns but differed from the labels in peak amplitude, timing, and baseline level. The relative importance of these error types varied across tasks and joint angles. The examples therefore provide information that is not visible from the overall MAE, RMSE, or Pearson \(r\) alone.

\subsubsection{Target Setting Comparison}
\label{subsubsec:target_setting_comparison}

To examine how the target time affects estimation performance, \textit{Random Forest} and \textit{TCN} were evaluated under the three target settings using both within subject and cross subject splits.
\begin{table*}[!t]
\caption{Results for \textit{Random Forest} and \textit{TCN} under the three target settings. MAE and RMSE are in degrees.}
\label{tab:alignment_results}
\centering
\scriptsize
\setlength{\tabcolsep}{3.5pt}
\renewcommand{\arraystretch}{1.08}
\begin{tabular*}{\textwidth}{
@{\extracolsep{\fill}}
l l l r r r}
\toprule
Target setting &
Evaluation &
Model &
MAE (\(\degunit\)) &
RMSE (\(\degunit\)) &
Pearson \(r\) \\
\midrule

\textit{Window-end}
& Within subject
& \textit{Random Forest}
& 9.457 & 13.183 & 0.678 \\

\textit{Window-end}
& Within subject
& \textit{TCN}
& 8.388 & 12.153 & 0.733 \\

\textit{Window-end}
& Cross subject
& \textit{Random Forest}
& 9.591 & 13.639 & 0.634 \\

\textit{Window-end}
& Cross subject
& \textit{TCN}
& 8.840 & 12.515 & 0.672 \\

\midrule

\textit{Centered-window offline}
& Within subject
& \textit{Random Forest}
& 9.109 & 12.711 & 0.703 \\

\textit{Centered-window offline}
& Within subject
& \textit{TCN}
& 7.897 & 11.385 & 0.766 \\

\textit{Centered-window offline}
& Cross subject
& \textit{Random Forest}
& 9.260 & 13.182 & 0.663 \\

\textit{Centered-window offline}
& Cross subject
& \textit{TCN}
& 8.613 & 12.215 & 0.698 \\

\midrule

\textit{Forecasting}
& Within subject
& \textit{Random Forest}
& 9.761 & 13.549 & 0.658 \\

\textit{Forecasting}
& Within subject
& \textit{TCN}
& 8.772 & 12.702 & 0.706 \\

\textit{Forecasting}
& Cross subject
& \textit{Random Forest}
& 9.864 & 13.969 & 0.614 \\

\textit{Forecasting}
& Cross subject
& \textit{TCN}
& 9.470 & 13.527 & 0.620 \\

\bottomrule
\end{tabular*}
\end{table*}
For both models and both evaluation splits, \textit{Centered-window offline estimation} produced lower errors than \textit{Window-end estimation}, whereas \textit{Forecasting} produced higher errors. Cross subject results were consistently less accurate than within subject results.

\subsection{Ablation Study}
\label{subsec:ablation_study}

The ablation experiments examine how frequency content, anatomical site coverage, modality, sensor density, and the number of training subjects affect joint angle estimation. All experiments use the same \textit{Random Forest} configuration and \textit{Window-end estimation}. Except for the frequency experiment, all analyses use the primary \textit{5--100\,Hz} AMG input. Frequency, anatomical site, sensor density, and training subject experiments use cross subject evaluation, while the modality experiment reports both within subject and cross subject results.

Where shown in Fig.~\ref{fig:dataset_value_ablations}, error bars give 95\% intervals from 10,000 subject resamples. For each resample, subjects from the corresponding test set were drawn with replacement and their mean MAE was recalculated. Full details are provided in SI Section~S12.

\subsubsection{Frequency Ablation}
\label{subsubsec:frequency_ablation}

This experiment examines how the frequency content of AMG affects joint angle estimation. Four signal representations were generated independently from the same DC corrected AMG data. The filtered conditions were applied to each complete trial before window extraction. All conditions used the same cross subject split, test windows, feature calculation, and \textit{Random Forest} configuration.

\begin{table}[!t]
\caption{Cross subject frequency ablation using \textit{Random Forest} and \textit{Window-end estimation}. MAE and RMSE are in degrees.}
\label{tab:frequency_band_ablation}
\centering
\footnotesize
\setlength{\tabcolsep}{3pt}
\renewcommand{\arraystretch}{1.08}
\begin{tabularx}{\columnwidth}{
@{}
>{\raggedright\arraybackslash}X
r r r
@{}}
\toprule
Signal representation &
MAE (\(\degunit\)) &
RMSE (\(\degunit\)) &
Pearson \(r\) \\
\midrule
\textit{Full band}              & 7.751  & 10.941 & 0.778 \\
\textit{20\,Hz low-pass}        & 7.405  & 10.338 & 0.794 \\
\textit{5--100\,Hz band-pass}   & 9.591  & 13.639 & 0.634 \\
\textit{100--760\,Hz band-pass} & 12.215 & 16.608 & 0.387 \\
\bottomrule
\end{tabularx}
\end{table}

The \textit{20\,Hz low-pass} condition produced the lowest error, while the \textit{100--760\,Hz band-pass} condition produced the highest error. This pattern shows that low frequency movement components provide substantial information for joint angle estimation. 
We use the \textit{5--100\,Hz band-pass} data as the default benchmark input in the following studies because it removes low-frequency kinematic artifacts while preserving the dynamic skin responses necessary for tracking muscle activities.

\begin{figure*}[!t]
\centering
\includegraphics[width=\textwidth]{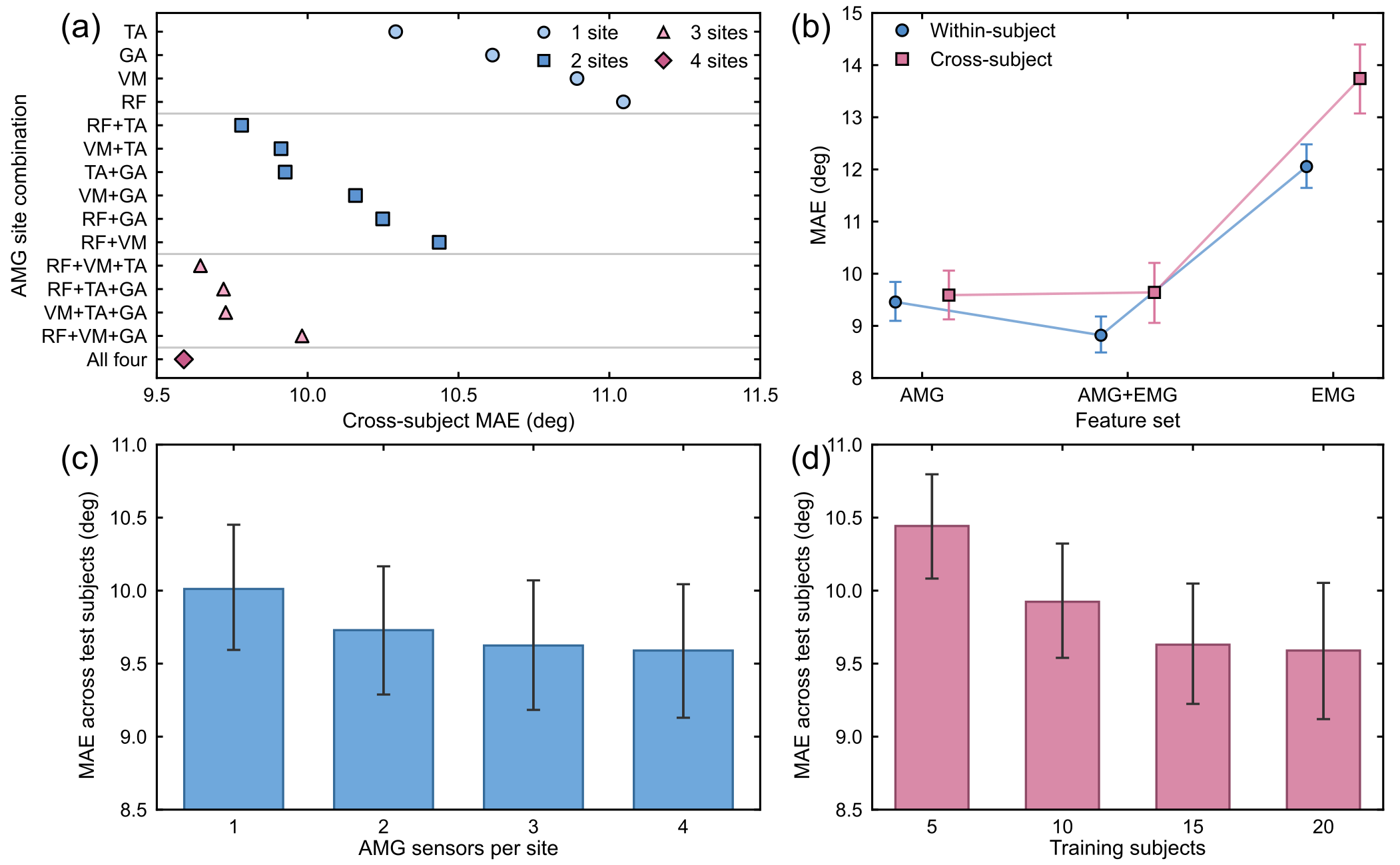}
\caption{Ablation results using the primary \textit{5--100\,Hz} AMG input, \textit{Random Forest}, and \textit{Window-end estimation}. MAE was first calculated within each test subject and then averaged equally across subjects. (a) Cross subject MAE for all 15 nonempty combinations of the four AMG sites. RF: rectus femoris; VM: vastus medialis; TA: tibialis anterior; GA: gastrocnemius. (b) Within subject and cross subject MAE using AMG, AMG+EMG, and EMG features. (c) Cross subject MAE using one to four AMG sensors at each site. (d) Cross subject MAE using 5, 10, 15, or 20 training subjects. Where shown, error bars give 95\% intervals obtained by resampling the test subjects.}
\label{fig:dataset_value_ablations}
\end{figure*}

\subsubsection{Anatomical Site Ablation}
\label{subsubsec:anatomical_site_ablation}

This experiment evaluates all 15 nonempty combinations of the rectus femoris, vastus medialis, tibialis anterior, and gastrocnemius AMG sites. Only the anatomical site combination was changed.
As shown in Fig.~\ref{fig:dataset_value_ablations}(a), the full four site configuration produced the lowest MAE. Configurations containing the same number of sites still showed different errors. The result suggests that both anatomical coverage and site selection affect joint angle estimation.

\subsubsection{Modality and Fusion Ablation}
\label{subsubsec:modality_fusion_ablation}

This experiment compares three feature sets: 384 AMG features, 16 EMG features, and their 400 dimensional combination. The same feature sets were evaluated with both within subject and cross subject splits.
Figure~\ref{fig:dataset_value_ablations}(b) shows that adding EMG features reduced the within subject MAE but did not reduce the cross subject MAE. EMG features alone produced the largest error in both evaluations. 

\subsubsection{Sensor Density Ablation}
\label{subsubsec:sensor_density_ablation}

This experiment retains all four anatomical sites and varies the number of AMG sensors used at each site from one to four. For a given density, the same local sensor positions were selected at all four sites.
Figure~\ref{fig:dataset_value_ablations}(c) shows that MAE decreased as the number of sensors at each site increased. The largest improvement occurred when increasing from one to two sensors at each site, while the improvement from three to four sensors was small. 
These results demonstrate that adding \gls{amg} sensors to same muscle site improves estimation performance, but the marginal gains decrease as sensor density increases.

\subsubsection{Training Subject Number Ablation}
\label{subsubsec:training_subject_ablation}

This experiment evaluates training sets containing 5, 10, 15, or 20 subjects while keeping the validation and test groups unchanged. Five training subsets were used for the 5, 10, and 15 subject conditions, while the 20 subject condition used the complete training set.
Figure~\ref{fig:dataset_value_ablations}(d) shows that MAE decreased as the number of training subjects increased. The improvement was largest between 5 and 10 training subjects and became small between 15 and 20 subjects. 
The results indicate that increasing subject diversity enhances cross-subject estimation accuracy, albeit with marginal gains as the training dataset expands.

\section{Discussion}

This dataset combines dense AMG and matched EMG with MoCap joint angles during functional actions and treadmill walking. 
The cross-subject benchmark demonstrates that kinematic-filtered \gls{amg} data remain effective for predicting lower-limb movement.
Variation across subjects, tasks, and joint angles shows that a single overall score does not fully describe performance.

The ablation results clarify how signal modalities and sensor design affect estimation. The lower errors obtained from the \textit{Full band} and \textit{20\,Hz low-pass} inputs show that low frequency whole limb motion contributes strongly to joint angle estimation. 
The \textit{5--100\,Hz band-pass} \gls{amg} input eliminate low-frequency components with minimal impact on performance, with an error increase of $\sim 3^\circ$. Although the \textit{100--760\,Hz band-pass} input yields a higher error penalty ($\sim 6^\circ$), it remains capable of tracking joint angle kinematics.

Combining \gls{emg} with \gls{amg} data improved within subject estimation but not cross-subject estimation under the settings defined in Section.~\ref{sec:dataset_benchmark} A. 
Broader anatomical coverage, higher sensor density, and more training subjects improved performance, although with smaller gains with increased quantities.

These results show how the dataset can support studies of joint angle estimation, multimodal fusion, and sensor configuration. 
Note that the reference models evaluated in our benchmark tests serve only as baseline indicators and do not represent the performance upper bound for body movement tracking using \gls{amg} and \gls{emg} data. Future work should explore more advanced architectures to enhance estimation accuracy.

\section{Limitations}
The current work has several limitations. First, the dataset was collected from 30 healthy adults during structured laboratory tasks. It does not include patients, clinical outcomes, free living activity, repeated measurements across days, long term wear, or sensor reattachment. Clinical use and deployment in daily environments require separate studies.

Secondly, AMG and EMG were recorded only from the left leg, while joint angle labels were available for both legs. Results for the right leg should therefore not be interpreted as direct sensing of right leg muscle activity. 
Joint angles were derived directly from optical MoCap markers and the corresponding segment vectors, rather than using a clinical joint coordinate system based on rigorous inverse kinematics. 
Future work should incorporate more accurate, full-body kinematic tracking to further validate these findings.

The multimodal data were aligned using recorded start times, nominal sample rates, interpolation, and a shared time range. No additional correction for clock offset or drift was applied. Because the AMG and EMG inputs were filtered over a 15-second trial with zero phase filters, the reported benchmark is based on an offline setting; for real time prediction, filtering a shorter data segment may decrease the estimation performance. 

\section{Conclusion}
This work presents a synchronized multimodal dataset of lower-limb movement, comprising 1,918 trials across 16 tasks from 30 subjects, with integrated AMG, EMG, and motion-capture joint angles.
Specifically, our dataset offers a unique perspective on tracking muscle activity via dynamic skin movements near muscle sites, highlighting AMG as a key mechanical alternative to traditional EMG. Benchmark evaluations demonstrate AMG's efficacy in capturing fine-grained muscle activity, achieving superior joint angle tracking at a lower hardware cost.
We also conduct an in-depth investigation into AMG usage through ablation studies, demonstrating how frequency content, sensor layout, recording modality, and the number of training subjects impact joint-tracking performance.
Overall, this dataset and its benchmark results establish a solid foundation for future research in joint angle estimation, multimodal fusion, sensor selection, and cross-subject generalization, offering valuable implications for rehabilitation and movement science.

\section{Data and Code Availability}

The anonymized dataset and accompanying code are publicly available at \url{https://doi.org/10.57967/hf/9950}. The repository includes the processed trial data, metadata, data dictionary, data split files, window definitions, and benchmark resources described in this paper.




\makeatletter
\renewenvironment{thebibliography}[1]
{%
  \section*{References}%
  \addcontentsline{toc}{section}{References}%
  \normalcolor
  \footnotesize
  \vskip 0.3\baselineskip\relax
  \list{\@biblabel{\@arabic\c@enumiv}}{%
    \settowidth{\labelwidth}{\@biblabel{#1}}%
    \setlength{\leftmargin}{\labelwidth}%
    \addtolength{\leftmargin}{\labelsep}%
    \setlength{\itemsep}{0pt}%
    \setlength{\parsep}{0pt}%
    \setlength{\topsep}{0pt}%
    \usecounter{enumiv}%
    \let\p@enumiv\@empty
    \renewcommand{\theenumiv}{\@arabic\c@enumiv}%
  }%
  \let\@IEEElatexbibitem\bibitem
  \def\bibitem{\@IEEEbibitemprefix\@IEEElatexbibitem}%
  \def\newblock{\hskip .11em plus .33em minus .07em}%
  \sloppy
  \clubpenalty4000
  \widowpenalty4000
  \interlinepenalty500
  \sfcode`\.=1000\relax
}
{%
  \def\@noitemerr{%
    \@latex@warning{Empty `thebibliography' environment}%
  }%
  \endlist
}
\makeatother

\bibliographystyle{IEEEtran}
\bibliography{dataset_refs/phase7_intro_references_ieee_ascii,dataset_refs/intro_expansion_verified_20260707}

\end{document}


\maketitle

\section{Scope of the Supplementary Information}

This Supplementary Information (SI) provides additional details of subject
preparation, data acquisition, trial timing, task instructions, signal
processing, data splits, model settings, and statistical analysis. These
details supplement the main manuscript and support reproduction of the
released dataset and benchmark.

Acceleromyography (AMG) refers to the acceleration recorded by the skin-mounted accelerometers at
the selected muscle sites and to the AMG inputs derived from these recordings.
For the primary benchmark, the \SIrange{5}{100}{Hz} band-pass representation
was used to remove low-frequency kinematic artifacts while preserving the
dynamic skin responses required for tracking muscle activities.

Supplementary Sections S2--S8 describe the subjects, acquisition system, task
protocol, auditory cues, treadmill walking, heart-rate monitoring, and data
organization. Section S9 describes signal processing, window construction, and
data splits. Section S10 describes the frequency conditions. Section S11
reports the reference model and training settings. Section S12 defines the
statistical aggregation and subject resampling procedures. Section S13
summarizes the relation between the released data and the benchmark. Section
S14 provides the joint-angle marker definitions and validity rules.

\section{Subjects and Pre-session Preparation}

The dataset was collected from 30 healthy adults, comprising 15 female and
15 male subjects. Cohort summary statistics and the handling of
subject-level demographic fields are reported in the main manuscript. The
study protocol was approved by the Medical Ethics Committee of Harbin
Institute of Technology (Approval No.\ HIT-2024046; July 8, 2024). Every subject provided written informed consent before data collection and
received compensation.

Before recording, study personnel explained the experimental sequence,
demonstrated the required movements, and allowed each subject to become
familiar with the tasks and treadmill. Recording began only after the
subject understood the instructions and could perform the required form.
The acquisition-side subject record included sex, age, height, weight,
exercise habits, and self-reported history of lower-limb injury. Treatment of
subject-level fields is defined by the data dictionary accompanying the
dataset and the applicable ethics, consent, de-identification, and access
controls.

\section{Multimodal Data-Collection System}

\subsection{System architecture}

The acquisition system comprised one Windows host and four Raspberry Pi 5
units. The Windows host ran the central control application, the four-channel
surface-EMG acquisition application, and the optical motion-capture
application. Each Raspberry Pi acquired one cluster of four triaxial
accelerometers. The device clocks were coordinated using the Network Time
Protocol (NTP), and the central application coordinated the acquisition
programs and transfer of their output files to a shared collection directory.

NTP coordination and centrally controlled program startup supported
acquisition management but do not establish hardware-level clock accuracy. The
retained timestamps allow the timing of each recorded stream to be inspected.
The benchmark uses trial-relative timing and applies no cross-device correction
for clock offset or drift.

\begin{table}[htbp]
\centering
\caption{Acquisition components and their roles.}
\label{tab:si_acquisition_components}
\small
\begin{tabularx}{\textwidth}{
  >{\raggedright\arraybackslash}p{0.16\textwidth}
  >{\raggedright\arraybackslash}p{0.22\textwidth}
  >{\raggedright\arraybackslash}X
  >{\raggedright\arraybackslash}p{0.17\textwidth}}
\toprule
Component & Platform/hardware & Recorded information & Nominal sampling \\
\midrule
Accelerometer-based AMG &
Four Raspberry Pi 5 units and 16 triaxial AIS2IHTR-based accelerometers &
Four muscle-site accelerometer clusters for AMG recording; four
accelerometers per cluster and 48 acceleration axes in total &
\SI{1600}{Hz} \\

Surface EMG &
Windows host and four DataLITE WS450 LE230 sensors &
Four channels positioned over the rectus femoris, vastus medialis,
tibialis anterior, and gastrocnemius &
\SI{1000}{Hz} \\

Optical motion capture &
Windows host, 16 Mars2H infrared cameras, and 15 reflective markers &
Marker trajectories used to construct bilateral knee and ankle geometric
joint-angle labels &
\SI{90}{Hz} \\

Heart-rate monitoring &
Huawei Band 9 &
Displayed heart rate used only for safety and recovery monitoring; no
heart-rate data were recorded or released &
Not applicable \\
\bottomrule
\end{tabularx}
\end{table}

\subsection{Muscle-site sensing layout}

Surface-EMG sensors were positioned over the rectus femoris, vastus medialis,
tibialis anterior, and gastrocnemius. At each site, four miniature triaxial
accelerometers were attached to the surrounding skin, forming one muscle-site
accelerometer cluster for AMG recording. Each cluster was connected to a
dedicated Raspberry Pi 5. This layout preserved a direct mapping among
anatomical site, accelerometer cluster, matched EMG channel, acquisition
stream, and dataset metadata. The 15 optical motion-capture markers were
placed at the lower-limb landmarks used to construct the geometric joint-angle
labels described in the main manuscript.

\subsection{Acquisition startup and storage}

For each trial, the motion-capture application was initiated first, followed
by the EMG and accelerometer acquisition applications under central control.
The applications produced modality-specific files with recorded timing
information. EMG and motion-capture output files were written to a shared
directory on the Windows host. Files generated by the four Raspberry Pi units
were transferred to the same host through mounted network directories.
Heart rate was viewed during collection but was not stored as a dataset
modality.

\section{Trial Organization and Auditory Cues}

Each subject was expected to complete four trials for each of the 16 task
conditions. The four trials were labeled A1, A2, A3, and A4. This design gives \(30\times16\times4=1{,}920\) expected trials. A total of 1,918 trials were retained. Two \emph{Walk at 2 km/h} trials were excluded because each was missing one required accelerometer-cluster file.

For the repeated functional actions, three auditory cues at approximately 2, 6, and 10 s were used to guide movement timing. Treadmill walking and the static posture tasks were not performed as repeated cue-paced movements. For \emph{Single-leg Stance}, the subject raised the right leg before recording, and the trial started at 0\,s after balance was stable.
The acquisition label ``RING'' identifies the auditory-cue records. Cue records were available for 1,726 of the 1,918 retained trials and were absent for 192 retained trials. Across all 1,920 expected trials, cue records were available for 1,728 trials and were absent for 192 trials. The two excluded trials both had cue records.
The released metadata state whether a cue record is available for each trial.
For the 1,726 retained trials with cue records, the metadata also provide the cue onset times relative to the trial start. Absolute cue timestamps are not released.
The auditory cues were used only to guide task execution. They were not used for multimodal alignment, joint-angle construction, window construction, data splitting, model training, or metric calculation.

Unless hand contact was required by the task, subjects kept their arms crossed over the chest during functional-action recording. Study personnel observed each trial and repeated the instructions when the task was not performed as specified.

\begin{figure*}[h]
\centering
\includegraphics[width=100mm]{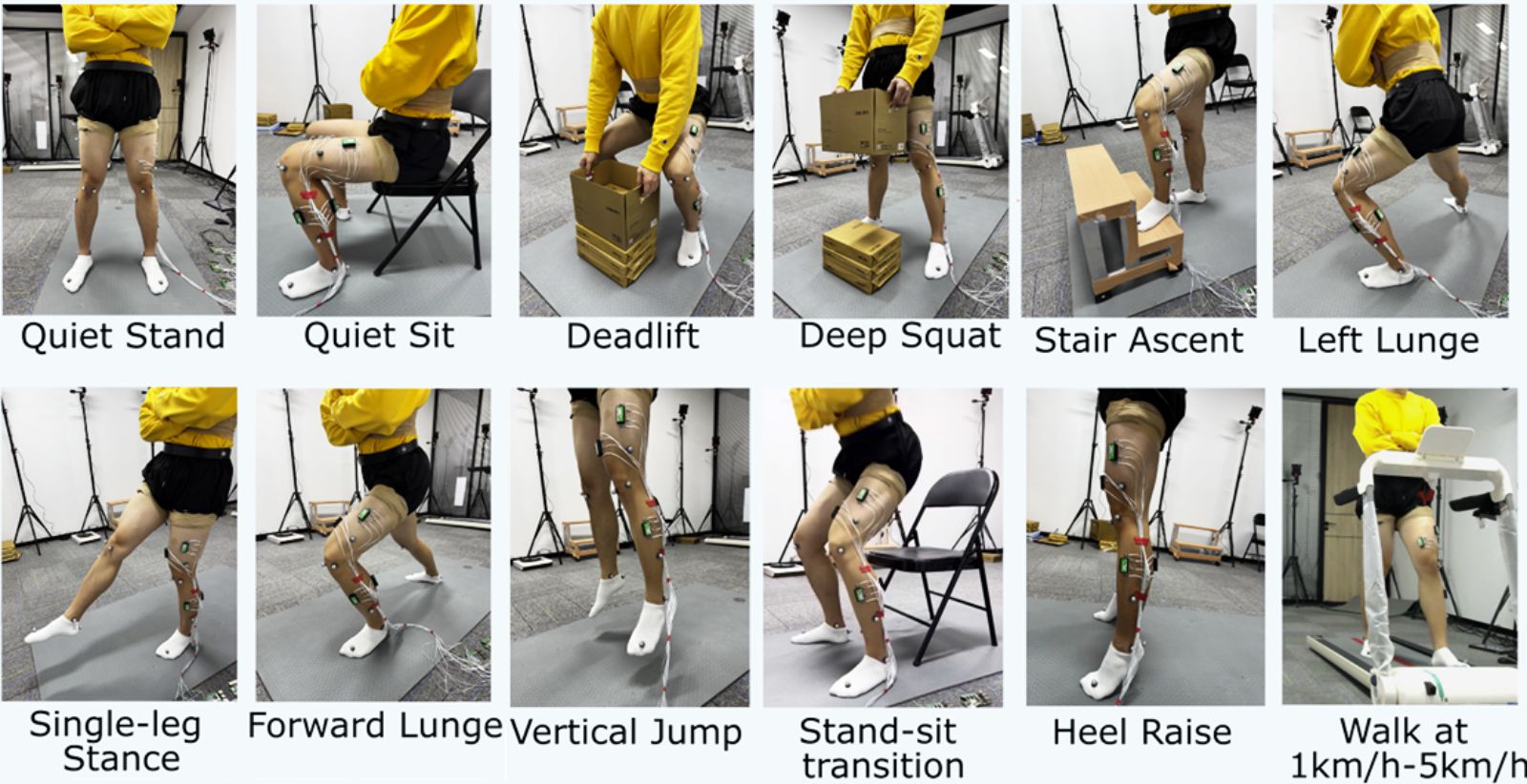}
\caption{The 11 functional lower limb actions and treadmill walking at 1--5 km/h.}
\label{fig:si_actions}
\end{figure*}

Table~\ref{tab:si_action_instructions} records the task instructions used to
standardize the 11 functional-action conditions. The task names match those
used in the main manuscript and data dictionary.
The photos of those actions are shown in Fig.~\ref{fig:si_actions}.
\section{Functional-Action Instructions}

\small
\begin{longtable}{
  >{\raggedright\arraybackslash}p{0.22\textwidth}
  >{\raggedright\arraybackslash}p{0.70\textwidth}}
\caption{Operational instructions for the functional-action conditions.}
\label{tab:si_action_instructions}\\
\toprule
Task condition & Standardized instruction \\
\midrule
\endfirsthead

\multicolumn{2}{l}{\small\upshape Table~\thetable\ continued.}\\
\toprule
Task condition & Standardized instruction \\
\midrule
\endhead

\midrule
\multicolumn{2}{r}{\small\upshape Continued on the next page.}\\
\endfoot

\bottomrule
\endlastfoot

Quiet Stand &
Stand upright with both feet approximately shoulder-width apart and maintain
the posture for the recording interval. \\

Quiet Sit &
Sit on the chair with the legs relaxed and the trunk upright, and maintain the
posture for the recording interval. \\

Deadlift &
Begin in standing with the feet approximately shoulder-width apart and the
toes turned slightly outward. The box contained two 500-sheet reams of A4
paper as the standardized load. The initial box height was adjusted using A4
paper reams placed beneath the box: one ream for the nominal 155--165-cm
height band, two reams for the nominal 165--175-cm band, and three reams for
subjects taller than approximately 175 cm. Study personnel selected the
appropriate recorded height band for each subject. The support-stack
corners were aligned with the directions of the feet. The subject lowered
and lifted the box within approximately \SI{4}{s}, kept the trunk upright
during the lifting phase, and looked forward. \\

Deep Squat &
Use the same initial foot placement and height-adjusted box arrangement as in
the Deadlift condition, but without the standardized load inside the box.
Complete the squat movement according to the auditory cues. \\

Stair Ascent &
Lead the upward phase with the left leg and use the right leg for the downward
phase, following the same step arrangement across trials. \\

Left Lunge &
Place the left foot at approximately \(45^\circ\), allow the right leg to
remain relaxed, and advance the left knee forward beyond the toes according to
the standardized study form. Complete the movement according to the auditory cues. \\

Single-leg Stance &
Raise the right leg before recording. After balance is stable, start the trial
at 0 s and maintain the posture for the full recording interval. Keep the toes
of the raised foot approximately parallel to the floor. \\

Forward Lunge &
Use a relative foot angle of approximately \(30^\circ\) and advance the
front knee forward beyond the toes according to the standardized study form.
Complete the movement according to the auditory cues. \\

Vertical Jump &
Begin with the feet approximately shoulder-width apart, perform a vertical
jump in place, and return to the starting area according to the auditory cues. \\

Stand-sit transition &
Position the chair so that its front edge is aligned with the posterior
lower-leg/heel line. Begin in standing and perform the instructed transition
according to the auditory cues. \\

Heel Raise &
Begin with both feet approximately shoulder-width apart and perform a
bilateral heel raise within each cue interval. \\

\end{longtable}
\normalsize

\section{Treadmill-Walking Conditions}

Treadmill walking was recorded at 1, 2, 3, 4, and \SI{5}{km/h}, giving five
conditions separated by \SI{1}{km/h} (Fig.~\ref{fig:si_actions}). Four trials were scheduled at each
speed. Before each recording, the treadmill was brought to the target speed
and the subject was allowed to adapt until walking appeared stable.
The approximately \SI{15}{s} trial was then recorded. Study personnel
continued to observe the subject, and the recording was paused when rest
was required.

\section{Heart-Rate Monitoring, Rest, and Fatigue Control}

Heart rate was viewed during data collection using a Huawei Band 9. It was used
only as a safety and recovery indicator. Recording was paused when the
displayed heart rate was elevated relative to the subject's resting or
recovery state, or when the subject reported fatigue or discomfort.

The subject rested before recording resumed. Recording continued only after
the displayed heart rate had decreased and the subject confirmed readiness to
continue. Heart-rate data were not recorded and are not included in the
released dataset. Heart rate was not used as a model input, joint-angle label,
or quantitative measure of fatigue.

\section{Acquisition-side Data Organization}

Acquisition files were organized hierarchically by subject, task
condition, and trial. Each subject had one acquisition-side record
containing the subject metadata collected for the study and corresponding
subdirectories containing the modality-specific trial files. The Windows
host's shared collection directory received the EMG, motion-capture, and four
accelerometer-stream files. No heart-rate data file is included in the processed dataset.

In the processed dataset, stable subject--task--trial identifiers link the
AMG, EMG, motion-capture, timing, annotation, and joint-angle-label contents.
Exact array names, dimensions, units, task-code mappings, excluded-trial
identifiers, split membership, and metadata-access designations are defined in
the versioned data dictionary rather than inferred from acquisition-side
directory names.

\section{Benchmark Preprocessing and Data Partitions}

\subsection{AMG correction and primary benchmark input}

Let \(x_c[n]\) denote sample \(n\) from aligned AMG channel \(c\), where
\(c=1,\ldots,48\) and \(n=0,\ldots,N_c-1\). Let \(N_c\geq1\) denote the
number of aligned samples in that channel for the current trial.DC correction was defined as
\begin{align}
n_0 &= \min(3200,N_c),\\
\bar{x}_{c,0} &= \frac{1}{n_0}\sum_{n=0}^{n_0-1}x_c[n],\\
x^{\mathrm{DC}}_c[n] &= x_c[n]-\bar{x}_{c,0}.
\end{align}
The offset was therefore calculated from the first 3,200 samples, equivalent
to the first \SI{2}{s} at \SI{1600}{Hz}. If a channel contained fewer than
3,200 samples, all available samples were used.

The primary AMG representation was generated by applying a Butterworth
\SIrange{5}{100}{Hz} band-pass filter to every complete DC-corrected trial and
channel before window extraction. The filter was designed using SciPy libarary, by setting an order parameter of 4. 
The filter used forward and backward application with odd extension and a padding length of 27 samples. In the main manuscript, this representation is referred to as the kinematic-filtered AMG input. The filtering was used to eliminate the
impact of body kinematics while preserving the skin dynamics used for
joint-angle estimation. This processing was offline and zero phase.

Let \(\mathbf{t}_{\mathrm{AMG}}\) denote the ascending common AMG time vector, and let \(t_{S,\mathrm{AMG}}\) denote its first timestamp. Let
\(j=0,1,\ldots\) index the candidate endpoint times. For candidate endpoint time \(a_j\), let \(e_j\) denote the exclusive stop index and \(s_j\) denote the inclusive start index:
\begin{align}
a_j
&=
t_{S,\mathrm{AMG}}
+
\left(0.300+0.100j\right)\,\mathrm{s},
\\
e_j
&=
\operatorname{searchsorted}
\left(
\mathbf{t}_{\mathrm{AMG}},
a_j,
\text{side}=\text{right}
\right),
\\
s_j
&=
e_j-480.
\end{align}
Here, the right-sided search returns the first index after all timestamps less than or equal to \(a_j\). The resulting window contains the samples with indices \(s_j,\ldots,e_j-1\), giving exactly 480 samples. Only candidate endpoints whose full 480-sample support lies within the common AMG time vector are retained.
Each AMG window therefore contained 480 samples from 48 channels and covered \SI{300}{ms}. The dataset contained 264,081 base windows before joint-angle label validity was checked.For each retained window \(i\) generated from candidate endpoint \(j\), the notation in the main manuscript satisfies \(n_i=s_j\) and \(t_i^{\mathrm{end}}=a_j\).

\subsection{EMG processing and feature representation}

EMG remained at its native sampling rate of \SI{1000}{Hz}. It was not
resampled to the AMG time grid. A Butterworth \SIrange{20}{450}{Hz} band-pass
filter was applied to each complete EMG trial using forward and backward
second-order-section filtering. The SciPy design used an order parameter of 4, four second-order sections, odd extension, and a padding length of 27 samples. The filtered EMG was not rectified, and no envelope was calculated.

Each AMG window was paired by time with a \SI{300}{ms} EMG window containing
300 native samples from each of the four EMG channels. Mean absolute value,
waveform length, zero crossings, and slope sign changes were calculated for
each channel. These four features from four channels gave 16 EMG features.
Combining them with the 384 AMG features gave a 400-dimensional feature
vector.

\subsection{Trial identifiers, data splits, and window counts}

A1, A2, A3, and A4 identify the four recorded trials for one subject and task
condition. Within subject evaluation used A1 and A2 for training, A3 for
validation, and A4 for testing. Trial assignment was completed before window
extraction.

Cross subject evaluation assigned 20 subjects to training, 3 subjects to
validation, and 7 subjects to testing. All trials from one subject remained
in the same partition. Subject identity was not used as a model input.

The complete valid-window counts are shown in
Table~\ref{tab:si_test_window_counts}. AMG and EMG were available for all
264,081 base windows. Differences among the three settings resulted from
joint-angle label validity.

\begin{table}[htbp]
\centering
\caption{Valid window counts for the three target settings and data splits.}
\label{tab:si_test_window_counts}
\small
\setlength{\tabcolsep}{3pt}
\begin{tabularx}{\textwidth}{
>{\raggedright\arraybackslash}X
rrrrrrr}
\toprule
Target setting &
All valid &
Within train &
Within validation &
Within test &
Cross train &
Cross validation &
Cross test \\
\midrule
Window-end estimation &
263,999 & 132,123 & 65,957 & 65,919 &
176,229 & 26,083 & 61,687 \\

Centered-window offline estimation &
263,994 & 132,119 & 65,955 & 65,920 &
176,225 & 26,083 & 61,686 \\

Forecasting &
263,833 & 132,036 & 65,920 & 65,877 &
176,129 & 26,062 & 61,642 \\
\bottomrule
\end{tabularx}
\end{table}

\section{Frequency Ablation}

The frequency experiment used cross subject
\emph{Window-end estimation} with \emph{Random Forest}. All four conditions
were generated from the same DC-corrected AMG trials. Each filter was applied
to the complete trial before window extraction. The same 61,687 cross subject
test windows and the same model settings were used for all conditions.

\begin{table}[htbp]
\centering
\caption{AMG signal representations used in the frequency experiment.}
\label{tab:si_frequency_conditions}
\small
\begin{tabularx}{\textwidth}{
>{\raggedright\arraybackslash}p{0.22\textwidth}
>{\raggedright\arraybackslash}p{0.34\textwidth}
>{\raggedright\arraybackslash}X}
\toprule
Signal representation & Filter design & Role \\
\midrule
Full band &
No additional frequency-selective filter applied to the DC-corrected signal &
Comparison \\

\SI{20}{Hz} low pass &
SciPy Butterworth filter designed with order parameter 4; two second-order sections;
forward and backward filtering; padding length 15 &
Comparison \\

\SIrange{5}{100}{Hz} band pass &
SciPy Butterworth filter designed with order parameter 4; four second-order sections; forward and backward filtering; padding length 27 &
Primary benchmark representation \\

\SIrange{100}{760}{Hz} band pass &
SciPy Butterworth filter designed with order parameter 4; four second-order sections; forward and backward filtering; padding length 27 &
Exploratory comparison \\
\bottomrule
\end{tabularx}
\end{table}

The \SIrange{5}{100}{Hz} band-pass representation was used as the primary
benchmark input because it removes low-frequency kinematic artifacts while
preserving the dynamic skin responses necessary for tracking muscle
activities. It was therefore used as the default AMG representation in the
subsequent benchmark and ablation experiments. All filtered conditions were
generated offline by filtering complete trials before window extraction.

\section{Reference-Model Implementations}

\subsection{Random Forest}

Random Forest was used as the reference model for the feature inputs. It used
100 trees, a maximum depth of 15, and random seed 42. The same configuration
was used in the primary feature benchmark and the frequency, modality,
anatomical-site, sensor-density, and training-subject experiments.

\subsection{Temporal architectures}

The temporal models were implemented using tsai version 1.0.1. Each model
received one \(48\times480\) AMG window and predicted four joint angles.
Table~\ref{tab:si_temporal_models} lists the model settings and evaluation
scope.

\begin{table}[htbp]
\centering
\caption{Temporal model settings and evaluation scope.}
\label{tab:si_temporal_models}
\small
\begin{tabularx}{\textwidth}{
>{\raggedright\arraybackslash}p{0.14\textwidth}
>{\raggedright\arraybackslash}p{0.24\textwidth}
>{\raggedright\arraybackslash}p{0.32\textwidth}
>{\raggedright\arraybackslash}X
r}
\toprule
Model & tsai class & Architecture & Evaluation scope & Seed \\
\midrule
TCN &
\path{tsai.models.TCN.TCN} &
Eight temporal convolution layers; 25 channels per layer; kernel size 7;
76,154 trainable parameters &
All three target settings; within subject and cross subject &
42 \\

LSTMPlus &
\path{tsai.models.RNNPlus.LSTMPlus} &
One unidirectional LSTM layer; hidden size 100; final-step regression head;
60,404 trainable parameters &
Cross subject \emph{Window-end estimation} &
42 \\

InceptionTimePlus &
\path{tsai.models.InceptionTimePlus.InceptionTimePlus} &
Residual depth 6; 32 filters; kernel size 40; bottleneck layers;
464,388 trainable parameters &
Cross subject \emph{Window-end estimation} &
42 \\
\bottomrule
\end{tabularx}
\end{table}

\subsection{Training configuration}

Each reported model configuration was trained once. Input normalization used
one mean and one population standard deviation for each AMG channel. Target
normalization used one mean and one population standard deviation for each
joint angle. All normalization values were calculated from the training
partition only.

Training used mean squared error loss, AdamW, a learning rate of \(10^{-3}\),
weight decay of \(10^{-4}\), gradient clipping at 1.0, and mixed precision.
Training was limited to 1,000 epochs. It stopped after 50 consecutive epochs
without a validation MAE improvement of at least \(10^{-4}\). The checkpoint
with the lowest validation MAE was retained.

TCN used a training batch size of 4,096 and an evaluation batch size of 8,192.
LSTMPlus and InceptionTimePlus used a physical training batch size of 512 and
gradient accumulation to obtain an effective batch size of 4,096. Their
evaluation batch size was 1,024. All runs used eight data-loader workers and
random seed 42. Test data were not used for checkpoint selection.

\section{Statistical Aggregation and Reference Prediction Reporting}

For the cross-subject benchmark, MAE, RMSE, and Pearson \(r\) were
calculated separately for the right knee, left knee, right ankle, and left
ankle over all valid test windows. The four joint-angle values were then
averaged arithmetically. For \emph{Window-end estimation}, the primary cross
subject test set contained 61,687 valid windows.

For subject analysis, each metric was calculated separately for every test
subject. For task analysis, MAE was first averaged over valid windows and the
four joint angles for each subject and task. The seven cross subject test
values were then averaged with equal weight. For joint-angle analysis, MAE was
calculated separately for each subject and each joint angle.
Where a 95\% interval is shown, subjects were resampled with replacement
10,000 times. Each resample contained the same number of subjects as the
corresponding test set. The mean MAE was recalculated for every resample, and
the 2.5th and 97.5th percentiles formed the interval. Windows were not treated
as independent resampling units.

For the modality experiment, one MAE was calculated for each subject and
feature set before subject resampling. Within subject intervals used the
within subject test subjects, and cross subject intervals used the seven cross
subject test subjects.

For the sensor-density experiment, results from all matched sensor subsets at
one density were first averaged within each test subject. The seven subject
values were then averaged, and the 95\% interval was obtained by resampling
subjects.

For the training-subject experiment, results from the predefined training
subsets of one size were first averaged within each test subject. The seven
subject values were then averaged, and the 95\% interval was obtained by
resampling subjects.

The anatomical-site panel reports the equal-subject mean for each site
combination and does not display error bars.

Each model table reports one reference prediction from the single training run
defined in Section S11. Figure 4 also uses one reference Random Forest
prediction and one reference TCN prediction. Predictions were not averaged
across repeated runs.

\section{Relationship to the Reported Benchmark}

The primary sequence input is the kinematic-filtered AMG representation
obtained by applying the \SIrange{5}{100}{Hz} band-pass filter to each complete
DC-corrected trial before window extraction. Each input window contains 48
channels and 480 samples per channel. The Random Forest AMG input contains 384
features. The EMG input contains 16 Hudgins time-domain features, and the
AMG+EMG input contains 400 features.

The four joint-angle outputs are stored in the following order: right knee,
left knee, right ankle, and left ankle. The angles are calculated from
marker-defined segment vectors. They are not clinical joint-coordinate
measurements or outputs from a full inverse-kinematics model.

Heart rate was not recorded as a dataset modality. Auditory-cue timing is
included in the released metadata but is not used as a model input, label,
alignment signal, or split variable. Subject identity and acquisition folder
names are also not used as model inputs.

The main manuscript reports the scientific results and conclusions. This SI
provides the processing, model, and statistical details needed to reproduce
the benchmark.

\section{Joint-Angle Construction Details}
\label{sec:supp_joint_angle_details}

The main manuscript defines the geometric knee and ankle angle operators and
their vector directions. Table~\ref{tab:supp_joint_angle_markers} records the
frozen marker order used by the released implementation. These indices are
implementation metadata rather than anatomical claims. All indices are
zero-based positions in the frozen marker array used by the released
implementation.
The MoCap system used in this work include 16 infrared motion capture cameras (Mars2H, NOKOV). The marker definitions are consistent with the labels used in the software: NOKOV XINGYING 3.4.0.3957, as listed in Table.~\ref{tab:supp_joint_angle_markers}. 
\begin{table}[htbp]
\caption{Frozen marker definitions used for the four geometric joint-angle outputs.}
\label{tab:supp_joint_angle_markers}
\centering
\small
\setlength{\tabcolsep}{4pt}
\renewcommand{\arraystretch}{1.10}
\begin{tabularx}{\textwidth}{
>{\raggedright\arraybackslash}p{0.15\textwidth}
>{\raggedright\arraybackslash}p{0.28\textwidth}
>{\raggedright\arraybackslash}X}
\toprule
Output & Frozen markers and indices & Vector directions \\
\midrule
Right knee &
R.Thigh (9), R.Knee (10), R.Ankle (12) &
R.Knee to R.Thigh; R.Knee to R.Ankle \\
Left knee &
L.Thigh (0), L.Knee (1), L.Ankle (3) &
L.Knee to L.Thigh; L.Knee to L.Ankle \\
Right ankle &
R.Shank (11), R.Ankle (12), R.Toe (14) &
R.Ankle to R.Shank; R.Ankle to R.Toe \\
Left ankle &
L.Shank (2), L.Ankle (3), L.Toe (5) &
L.Ankle to L.Shank; L.Ankle to L.Toe \\
\bottomrule
\end{tabularx}
\end{table}

Any marker-coordinate value with absolute magnitude greater than \(9\times10^{5}\) is set to missing. Nonfinite coordinates and segment vectors with Euclidean norms below \(10^{-8}\) propagate to missing labels. For two valid segment vectors \(\mathbf{u}\) and \(\mathbf{v}\), the normalized dot product
\(
\frac{
\mathbf{u}^{\mathsf{T}}\mathbf{v}
}{
\|\mathbf{u}\|_2\|\mathbf{v}\|_2
}
\) is clipped to \([-1,1]\) before arccosine evaluation. Missing labels are not replaced with zeros.

For benchmark label assignment, the MoCap frame nearest to the requested label time was used. If duplicate MoCap timestamps produced a tie, the earliest frame in the original order was selected. A window label was retained only when the requested time was within the valid MoCap time range, the nearest frame differed from that time by no more than \(1/90\) s, and all four joint angles at that frame were finite. Joint-angle labels were not interpolated or extrapolated.